\documentclass{article}
\usepackage{url}
\usepackage{latexsym}
\usepackage{amsmath}
\usepackage{graphicx}
\usepackage{amssymb}
\usepackage{algorithm}
\usepackage{algorithmic}
\usepackage{mathrsfs}
\usepackage{amstext}
\usepackage{amsgen}
\usepackage{amsthm}
\usepackage{array}
\usepackage{tabularx}
\usepackage{supertabular}
\usepackage[babel=true]{csquotes}

\usepackage{graphics}
\usepackage{graphicx}

\usepackage{amssymb}
\usepackage{amsthm}

\begin{document}

\title{Dependability of Sensor Networks for Industrial Prognostics and Health Management}

\author{Wiem Elghazel, Jacques M. Bahi, Christophe Guyeux,\\ Mourad Hakem, Kamal Medjaher, and Noureddine Zerhouni}

\maketitle

\begin{abstract}
Maintenance is an important activity in industry. It is performed either to revive a machine/component or to prevent it from breaking down. Different strategies have evolved through time, bringing maintenance to its current state: condition-based and predictive maintenances. This evolution was due to the increasing demand of reliability in industry. The key process of condition-based and predictive maintenances is prognostics and health management, and it is a tool to predict the remaining useful life of engineering assets. Nowadays, plants are required to avoid shutdowns while offering safety and reliability. Nevertheless, planning a maintenance activity requires accurate information about the system/component health state. Such information is usually gathered by means of independent sensor nodes. In this study, we consider the case where the nodes are interconnected and form a wireless sensor network. As far as we know, no research work has considered such a case of study for prognostics. Regarding the importance of data accuracy, a good prognostics requires reliable sources of information. This is why, in this paper, we will first discuss the dependability of wireless sensor networks, and then present a state of the art in prognostic and health management activities. 

\end{abstract}

\section{Introduction}
\label{Intro}

\indent During their life cycle, industrial systems are subject to failures, which can be irreversible or have undesirable outcomes with consequences varying from minor to severe. From this context, it is important to monitor a system, assess its health, and plan maintenance activities. Thus, it will be possible to avoid \enquote{catastrophic} failure results.\\
\indent Over the past years, research in Prognostic and Health Management (PHM) field has gained a great deal of attention. Prognostic models are developed in an attempt to predict the Remaining Useful Life (RUL) of machinery before failure takes place. A maintenance schedule is then decided and system shutdown is prevented. Yet, if the prediction model and the provided measurements are not accurate, it is possible that the maintenance activity will be performed either \enquote{too soon} or \enquote{too late}.\\
\indent Such a prediction activity requires online measurements of the operating conditions of the system under consideration. This information is usually gathered by means of sensor nodes. In this study, we consider the case where the nodes communicate their information within a Wireless Sensor Network (WSN). Nevertheless, a WSN is prone to failure due to the nature of communication in the network and to the characteristics of its devices. For this reason, before deployment, a prior dependability study of the network is needed. It is the only way to guarantee the reception of accurate data.\\
\indent Although both dependability of WSNs and prognostic models development have been studied and reported in the literature, As far as we know, none of the existing research work has considered the dependability of WSNs for PHM purposes. When it comes to prognostics, it is usually assumed that the available data is accurate and complete. These assumptions are far from reality and the resulting prognostic model cannot provide good results in real-life applications. Considering the limited computational capacities of WSNs, it is very common to privilege some dependability matters over others, regarding the target application's requirements. Thus, it is crucial to consider a \enquote{prognostic-oriented} dependability solution for WSNs.\\
\indent This paper presents dependability issues of WSNs that are relevant for RUL prediction and discusses different prognostic approaches. The remainder of the paper is structured as follows. Section 2 presents an overview of wireless sensor networks. A state of the art in prognostics and health management is provided in Section 3. The relation between prognostics and WSN dependability is illustrated in Section 4. Section 5 discusses the remaining challenges. Finally, a conclusion is given in Section 6.

\section{Overview of Wireless Sensor Networks}

\begin{figure}[!h]
\centering
\includegraphics[width=0.725\textwidth,height=0.225\textheight]{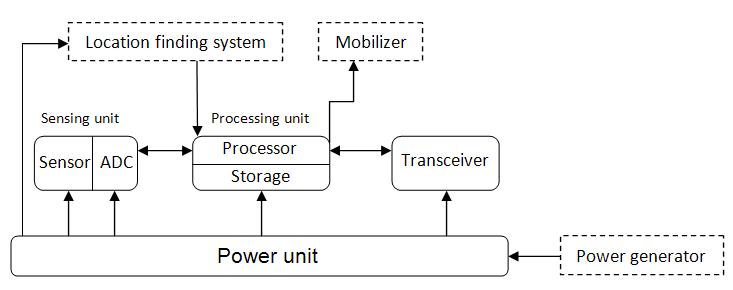}
\caption{Components of a sensor node}
\label{sensor}
\end{figure}

\indent WSNs are event-based systems that rely on the collective effort of several microsensor nodes \cite{Akan05}. This offers the network greater accuracy, larger coverage area, and the possibility to extract localized features. The network extends the computation capability to physical environments that human beings cannot reach. A sensor node is a tiny device having the capability of sensing new events, computing the sensed and received values, and communicating information. Thus, the network can be deployed to monitor physical and environmental phenomena such as temperature, vibrations, light, humidity, etc.\\
\indent Typically, a WSN is composed of few base stations and hundreds (or thousands) of sensor nodes. As shown in Figure \ref{sensor}, the nodes are equipped with sensors, data processing unit, memory space, radio communication range, and a battery.\\

There are different settings for a WSN model, which is generally dynamic, as radio range and network connectivity evolve over time \cite{Li11}. A network model can be either hierarchical, distributed, centralized, heterogeneous, or homogeneous \cite{Li11}.

The protocol stack of sensor networks is composed of five different layers \cite{Akyildiz02}.

\begin{enumerate}

\item \textbf{Physical layer:} it defines the means of transmitting data packets after being converted into raw data bits. Mainly, this layer is responsible for frequency selection, carrier frequency generation, signal detection, modulation, and data encryption.
\item \textbf{Data link layer:} it is charged of the creation of the network infrastructure, transferring data, and fairly and efficiently sharing the communication resources between sensor nodes. It is also responsible for error control of transmission data.
\item \textbf{Network layer:} it ensures internetworking with external networks, where the sink node can be used as a gateway. It commands and controls the system, forwards data packets, and takes charge of routing between intermediate routers.
\item \textbf{Transport layer:} it provides an end-to-end communication service, among other services such as multiplexing, reliability, flow control, congestion avoidance...
\item \textbf{Application layer:} it can be defined as the user interface. It displays messages in a human recognizable and understandable format.

\end{enumerate}

\subsection{Shortcomings of a WSN}

\indent WSNs are designed for the purpose of an efficient event detection. They consist of a large number of sensor nodes deployed in a surveillance area to detect  the occurrence of possible events. Such an activity necessitates efficiency, which is hard to achieve with the constraints of WSNs. These limitations are detailed in the following.

\subsubsection{Resources}

\indent Available energy is a big limitation to WSN capabilities. In fact, sensor nodes are small sized devices, resulting in tiny batteries as energy supply. Moreover, sensors are often deployed in hostile environments (mountains, enemy territory...) so once deployed, the sensor nodes cannot be recharged \cite{Carman00}.\\
\indent The added security code has an important impact on the available energy for normal network tasks. Processing security functions (encryption, decryption, signing data, verifications), transmitting security related data (vectors for encryption/decryption), and securing storage (cryptographic key) necessitate extra power, which is critical for WSNs \cite{Carman00, Walters07}.\\
\indent In addition to this, the deployed memory space for a sensor node is very limited. The storage space is shared between communication protocol and security code. The size of the latter has then to be limited to a minimum \cite{Walters07}.\\
Buffering space in sensor nodes is also limited. This will lead to packet loss with the increase in traffic flow towards the sink node. In fact, the area around the sink tends to be quickly congested as all sensor nodes tend to forward the captured data to the sink.

\subsubsection{Communication}

\indent Wireless communication is unreliable and this causes the network to be vulnerable. The absence of physical connections renders packet loss highly probable. Channel errors, missing links, route updates, network congestion can all cause packets drop.\\
\indent Multi-hop routing, network congestion, and node processing lead to a great latency and transmission errors in the network, and this makes synchronization among nodes hard to achieve. Synchronization is crucial for security matters (event reports, key distribution...) \cite{Walters07}.

\subsubsection{Operations}

\indent Harsh environment conditions and exposure to adversary attacks emphasize the likelihood of physical attacks on WSNs. Such attacks can cause permanent (even irreversible) damage to the hardware. Thus, the network will remain unable to fulfill the intended tasks \cite{Walters07}.\\
\indent Since the network is managed remotely, the sensor nodes are left unattended for a long period. It is therefore impossible to detect physical tampering and to perform regular maintenance.\\
\indent Routing solutions in WSNs avoid central management point as it may result in single point failure. Nevertheless, this solution may create an organization difficulty, leading to packet loss.

\subsubsection{Coverage and lifetime optimization}

\indent Considering all the limitations mentioned above, it is not easy for the network to always fulfill the intended tasks. Reliability and efficiency of WSNs are dependent on key issues, which are enumerated in the following.

\begin{enumerate}

\item Coverage

\indent Sensor nodes have a short radio range and they collaborate to cover a given surveillance area. At the network setup phase, it is crucial to ensure that the network is configured in a way that it covers all the area \cite{Tian05}. The coverage problem arises as: how to ensure that, at any time, any zone in the network is covered by at least one sensor node?\\
Zorbas \textit{et al.} \cite{Zorbas07} presented B\{GOP\}, a centralized coverage algorithm for WSNs. The algorithm proposes sensor candidate and avoids double-coverage depending on the coverage status of the corresponding field.\\
In \cite{Wang03}, Wang \textit{et al.} presented a protocol that can dynamically configure a network to achieve guaranteed degrees of coverage and connectivity. They gave a proof that sensing coverage range does not need to be more than half the connectivity range in the network. Thus, their protocol helps preserve energy while maintaining coverage in the network.\\
In \cite{HefeedaA10}, a general coverage algorithm, which considers the network connectivity is presented. The  proposed protocol, called Probabilistic Coverage Protocol (PCP), works for the common disk sensing model as well as probabilistic sensing model. To support probabilistic sensing models, the authors introduce the
notion of probabilistic coverage of a target area with a given threshold $\theta$, which means that an area is considered
covered if the probability of sensing an event occurring at any point in the area is at least $\theta$. They prove the correctness of the protocol and provide bounds on its convergence time and message complexity.\\

\item Awake nodes \textit{vs} sleeping nodes

\indent In order to prolong the network's lifetime, a possible solution is to keep a minimum number of sensor nodes in active mode. As WSNs rely on nodes density in the sensing and communicating processes, it is very likely that some nodes will not be needed. If a reliable node can forward data packets toward the sink, its neighbors can switch to idle state temporarily.\\
Lifetime optimization using knowledge about the dynamics of stochastic events has been studied in \cite{HeCLSS12}. The authors presented the interactions between periodic scheduling and coordinated sleep for both synchronous and asynchronous dense static sensor network. They show that the event dynamics can be exploited for significant energy savings, by putting the sensors on a periodic on/off schedule.\\
In \cite{HeCYSS12}, the authors design a polynomial-time distributed algorithm for maximizing the lifetime of the network. They proved that the lifetime attained by their algorithm approximates the maximum possible lifetime within a logarithmic approximation factor.\\
The authors in \cite{KasbekarBS11} leverage prediction to prolong the network life time, by exploiting temporal-spatial correlations among the data sensed by different sensor nodes. Based on Gaussian Process, the authors formulate the issue as a minimum weight submodular set cover problem and propose a centralized and a distributed truncated greedy algorithms (TGA and DTGA). They prove that these algorithms obtain the same set cover. \\

\item Transmission of wake-up messages

\indent As sensor nodes periodically go to sleep, they need to be awake when they are requested to. This is done by the transmission of wake-up messages towards a target sensor. However, if the message is not received at the right moment, data packets will be dropped. This will cost the network extra energy due to packet retransmission \cite{Ye03, Gallais06,Bahi11}.\\

\item Wear-out effect

\indent
In WSN, if the wear-out failures are not taken into consideration during the execution of the involved application, some nodes may age much faster than the others and become the reliability bottleneck for the network, thus significantly reducing the system's service lifetime. In the literature, this problem has been formulated and studied in various ways. For instance, prior work \cite{HeCLSS12,HeCYSS12,KasbekarBS11} in lifetime reliability  assumes node's failure rates to be independent of their usage times. While this assumption can be accepted for memoryless soft failures, it is obviously inaccurate for the wear-out-related fail-silent (a faulty node does not produce any output) and fail-stop (no node recovery) failures, because the sensor node's lifetime reliability will gradually decrease over time.\\
To cope with this problem, a distributed self-stabilizing and wear-out-aware algorithm is presented in \cite{BahiHHK2013}. This algorithm seeks to build resiliency by maintaining a necessary set of working nodes and replacing failed ones when needed. The proposed protocol is able to increase the lifetime of wireless sensor networks, especially when the reliabilities of sensor nodes are expected to decrease due to use and wear-out effects.

\end{enumerate}

\subsection{Attacks in WSNs}

\indent As discussed before, WSNs suffer from limited computation capabilities, a small memory capacity, poor energy resources, absence of infrastructure, and susceptibility to physical capture. A variety of security solutions exists for infrastructureless networks (Ad hoc networks). Yet, they do not all answer the security challenges of WSNs.\\
WSNs are vulnerable to many attacks, due to their uncontrolled environment of deployment, the limitation of their resources, and the broadcast nature of transmission medium. The attacks are mainly classified under two categories: physical attacks and non-physical attacks. In the following, we discuss some of the famous possible attacks in WSNs.

\subsubsection{Non-physical attacks}

\indent These attacks aim for disturbing the service and the normal operations in the network. In the following, examples of well-known non-physical attacks in WSNs are given.

\begin{enumerate}

\item Denial of Service (DoS) attack

\indent A DoS attack is an attempt to render the network unavailable to its users by leading to a server overload. It forces the network to either reboot or exhaust its resources, making it fail to perform the intended tasks. This attack is dangerous as WSNs lack the capacity to handle computational overhead to implement typical defensive strategies.\\
\indent This attack can target for physical layer (jamming, tampering), link layer (collisions, exhaustion, unfairness), network and routing layers (neglection, greed, homing, misdirection, black holes), or transport layer (flooding, desynchronization) \cite{Walters07}. A good review of DoS attacks in WSNs is given by Wood and Stankovic in \cite{Wood02}.\\
\indent DoS attacks are very common. Therefore, there exist effective defensive measures against these attacks. Table \ref{dos} summarizes the possible defenses against different DoS attacks.

\begin{table}[!h]
\centering
\small
\begin{tabular}{|c|l|l|}
\hline
Network Layer & Possible Attacks & Defenses \\
\hline
\hline
Physical Layer & - Jamming & - Spread-spectrum\\
&& - Priority messages \\
&& - Lower duty cycle\\
&& - Region mapping \\
&& - Mode change \\
\cline{2-3}
& - Tampering & - Tamper-proof \\
&& - Hiding \\
\hline
Link Layer & - Collision & - Error correcting codes \\
\cline{2-3}
& - Exhaustion & - Rate limitation \\
\cline{2-3}
& - Unfairness & - Small frames \\
\hline
Routing Layer & - Neglect and Greed & - Redundancy \\
&& - Probing \\
\cline{2-3}
& - Homing & - Encryption \\
\hline
& - Misdirection & - Egress filtering \\
&& - Authorization \\
&& -  Monitoring \\
\cline{2-3}
& - Black holes & - Authorization \\
&& - Monitoring \\
&& - Redundancy \\
\hline
Transport Layer & - Flooding & - Client puzzles\\
& - Desynchronization & - Authentication \\
\hline
\end{tabular}
\caption{DoS attacks for different network layers}
\label{dos}
\end{table}

Kim \textit{et al.} \cite{Kim06} proposed a DoS detection method reflecting the resource constraints of sensors. Their approach relied on two types of entropy estimators. A main estimator is charged of the synthesis of localized computations, whereas the other estimators are deployed hierarchically according to the network topology.\\
\indent Wood and Stankovic \cite{Wood02} described an approach to defend against jamming. In a first step, the nodes surrounding the jammed area report their status to they neighbors. In a second step, the neighbors collaborate to identify the jammed region so they will not route packets through it.\\

\item Sybil attack
\label{sybil}

\indent The Sybil attack was defined by Douceur in \cite{Douceur02} as \enquote{an attack able to defeat the redundancy mechanisms of distributed data storage systems in peer-to-peer networks}.\\
\indent A malicious device illegitimately takes on multiple identities to gain a large influence on the communication mechanism, by appearing and functioning as multiple distinct nodes. This attack is effective against routing algorithms, data aggregation, voting, fair resource allocation, and foiling misbehavior detection \cite{Walters07}.\\
\indent The Sybil attack is a harmful threat to sensor networks as it defeats redundancy mechanisms. To defend against such attacks, the network needs a mechanism to ensure that an identity is being held by one, and only one node in the network. A light-weight identity certificate method is proposed by Zhang \textit{et al.} \cite{Zhang05}. This method avoids public key cryptography by using one-way key chains and Merkle Hash trees.\\

\item Traffic analysis attack

\indent Even with encrypted data, traffic analysis is able to determine what type of information is being communicated in the network (chat, requests...) and even identify and disable the base station \cite{Walters07}. A malicious node can generate a fake physical event to be sensed. The attacker, then, will follow the data packets circulating in the network; nodes tending to forward more data packets are closer to the base station.\\
\indent Deng \textit{et al.} investigate three techniques that aim at hiding the true location of the base station \cite{Deng04}. The first technique is multi-parent  routing scheme. It introduces a degree of randomness in the multi-hop path from the source node until the base station. The second technique, called random walk, generates random fake routes to mislead an adversary while tracking the packet. And finally, fractal propagation consists in randomly creating multiple areas of high communication activities to disguise the true location of the base station.\\

\item Node replication attack

\indent In a node replication attack, an adversary can capture a sensor node, copy its ID, and insert a replica in the network \cite{Parno05}.\\
The consequences of such an attack can be severe; packets can be misrouted, changed, or corrupted, significant parts of the network can be disconnected \cite{Walters07}, etc.\\
\indent Parno \textit{et al.} argue that previous node detection replication schemes depend either on centralized mechanisms or on neighboring voting protocols \cite{Parno05}. Thus, they suffer either from single failure points or from failure to detect distributed replications. Considering these limitations, they propose two algorithms: randomized multicast and line-selected multicast. The line-selected multicast algorithm is inspired from the rumor routing described by Braginsky and Estrin in \cite{Braginsky02}. This algorithm helps reduce the communication cost of the randomized multicast protocol. 

\end{enumerate}

\subsubsection{Physical attacks}

\indent In a WSN, adversary can perform the following physical attacks:

\begin{itemize}
\item Known-Plaintext Attack (KPA): the attacker, having samples of both the plain text and the corresponding encryption, can reduce the security of the encryption key and reveal some of the information circulating in the network.
\item Chosen-Plaintext Attack (CPA): the attacker can choose a text to be encrypted. Doing so, it is easy to gain further information about the encryption key.
\item Man-In-The-Middle Attack(MITMA): the attacker creates a link between two nodes, through which they will communicate. The network cannot identify this connection as a malicious link. The attacker is then able to control the communication by intercepting and injecting messages.
\end{itemize}

\subsection{Dependability of WSNs}

\indent The dependability of a WSN is a property that integrates the attributes needed for the application to be justifiably trusted. Such a network should be able to deliver a correct service, i.e., a service that implements the system function, and makes sure that a failed component will not lead to system failure. System dependability was defined by Avizienis in \cite{Avizienis00} as \enquote{the ability of a system to avoid failures that are more frequent or more severe, and outage durations that are longer, than is acceptable to the users}.

\subsubsection{Threats}

\indent Developing a dependable WSN starts with defining the dependability requirements of users. In order to satisfy these needs, it is crucial to understand what might stop the network from delivering a correct service. In this section, the threats that can affect the dependability of a WSN are enumerated.

\begin{enumerate}

\item Faults

\indent A fault is the cause of an error, and it indicates a defect in a system. Its presence does not systematically lead to a failure. A fault is considered to be active only when it produces an error of one or more components. It is considered as transient if it affects the communication links between the nodes, and permanent if it is caused by hardware malfunction \cite{Silva12}.\\
\indent In \cite{Souza07}, the authors classify the sources of faults under two main categories: node faults (related to hardware) and network faults (related to routing).
A fault can be the result of various origins, which are classified in Figure \ref{fault}.\\

\begin{figure}[!h]
\centering
\includegraphics[width=0.7\textwidth,height=0.325\textheight]{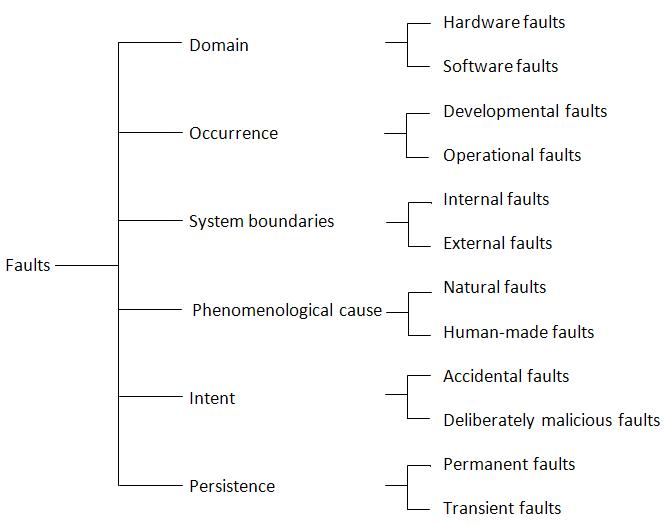}
\caption{Fault classes}
\label{fault}
\end{figure}

\item Errors

\indent An error takes place at runtime when some parts of the network enter in an unexpected (invalid) state that might result in a subsequent failure or undesirable outcomes. Such states are called hazards. As errors are hard to observe, special tools, such as debuggers, are required to declare their presence. An error indicates a discrepancy between actual behavior and intended behavior inside the network. It can then be detected if an error message or signal indicate its presence. If not detected, the error is called latent error \cite{Avizienis00, Taherkordi06}.\\

\item Failures

\indent
\indent A network failure is the observable consequence of an error. It occurs when the delivered service is no longer correct. The opposite transition (from incorrect to correct service) is called network restoration. Yet, the alteration of the service is not considered as a network failure until it reaches the service interface. With the implementation of fault tolerance techniques, a network failure can be avoided even when an error is activated \cite{Avizienis00, Silva12}. \\
The possible failure modes are outlined in Figure \ref{modes}.

\begin{figure}[!h]
\centering
\includegraphics[width=0.625\textwidth,height=0.25\textheight]{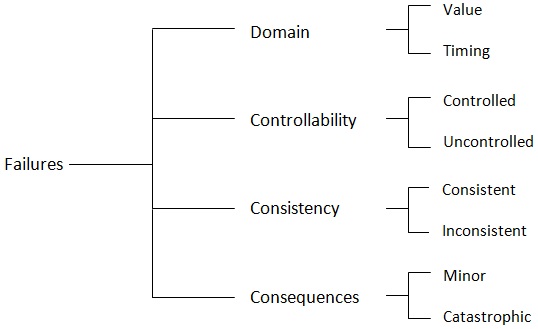}
\caption{Failure modes}
\label{modes}
\end{figure}

\end{enumerate}

\subsubsection{Attributes}

\indent This section deals with the ways by which we assess the dependability of a system. The attributes of dependability can vary in number and degree of importance considering the nature of the application and the intended service. The network, thus, is made dependable by adjusting the balance of the techniques to be employed according to the user's needs.

\begin{enumerate}















\item Availability

\indent In the classical definition, a network is considered as highly available if its downtime is very limited. This can be due either to few failures, or to quick restarts when failures take place \cite{Knight04, Taherkordi06}. If we add the security aspect, we can define availability as readiness for correct service for authorized users. This attribute can be computed as the probability that the network is functioning at a given time \cite{Silva12}.\\

\item Reliability

\indent A reliable network is a network that is able to continuously deliver a correct service. It can also be defined as the probability that a network functions properly and continuously in a time interval \cite{Silva12, Taherkordi06}.\\
\indent Most of research works that have been accomplished so far employ retransmission mechanisms over redundancy schemes to achieve network reliability. The main purpose of a WSN is the correct delivery of data packets from sensor nodes to end user. Thus, reliability of WSNs is highly related to data transport. Reliability can be classified into different levels: packet reliability, event reliability, Hop-by-Hop reliability, and End-to-End reliability.\\
\indent Both packet and event reliability levels deal with the required amount of information to notify the sink of the occurrence of an event within the network environment. Whereas the remaining two levels (i.e., Hop-by-Hop and End-to-End reliability levels) are concerned with the successful recovery of event information. Yet, all of them rely on retransmission and redundancy mechanisms.\\

\begin{enumerate}

\item Retransmission-based reliability

\indent Packet retransmissions is a very common technique to recover the loss of packets that did not arrive at their destination. This is generally ensured through the use of acknowledgements.\\
There mainly exist three different acknowledgement mechanisms that a receiver can employ to notify the sender of the reception status. The reception of an explicit acknowledgement (eACK) is a guarantee for the sender that the message was successfully delivered. In the opposite case, negative acknowledgement (nACK) means that the packet did not arrive at its intended destination. These two mechanisms increase the transmission overhead and thus consume much energy. Implicit acknowledgement (iACK) reduces energy consumption and it requires the sender to take benefit from the broadcast mechanism by listening to the channel and interpret the reception of the packet.\\
\indent Akan and Akyildiz presented the first event-based end-to-end reliability protocol \cite{Akan05}. This algorithm, which is called Event-to-Sink Reliable Transport (ESRT), has the ability of self-configuring according to the network condition, thus it is robust in dynamic network topologies.
\indent Zhou \textit{et al.} \cite{Zhou05} see reliability from a different angle compared to what is published in \cite{ Akan05, Gungor06, Iyer05}. They consider that reliability cannot just be measured by the total of incoming packets at the sink node. It should, instead, refer to nodes contribution to improve the sink information about a certain phenomenon. Price-Oriented Reliable Transport (PORT) protocol \cite{Zhou05} also considers an in-network congestion-avoidance mechanism as a remedy to the drawbacks related to end-to-end schemes. In the proposed scheme, data packets avoid the paths with high loss rates. As a result, PORT is more energy efficient compared to ESRT \cite{Akan05} and DST \cite{Gungor06}.\\

\item Redundancy-based reliability 
\label{redundancy}

\indent Reliability can also be introduced via data redundancy mechanisms. A packet is transmitted in multiple copies using different routes as a backup plan in case one route fails.\\
\indent Al-Wakeel and Al-Swailem \cite{Al-Wakeel07} proposed a Path Redundancy Based Security algorithm (PRSA) that defines secure multiple least cost routing paths between source and destination nodes. The optimum paths are selected referring to Dijkstra algorithm described in \cite{Dijkstra74}. When a node is suspected of being malicious, it will be removed from the routing path. Furthermore, PRSA allows source node to transmit data packets using various modes in order to enhance network security.\\
\indent Mojoodi \textit{et al.} \cite{Mojoodi11} studied the effect of redundancy on the number of correct responses of WSN on the received queries, and also investigate the change in the needed level of redundancy according to different network conditions. The simulations showed that redundancy is only efficient if the number of clusters needed to respond to the request is important or if error probability is high. In the opposite case, redundancy mechanisms will only lead to extra energy consumption and possible failures related to unnecessary communication cost \cite{Mojoodi11}.\\

\end{enumerate}

\item Security

\indent
\indent WSNs are different from traditional computer networks. Therefore, existing security mechanisms are not suitable for these networks. Developing adequate security measures requires understanding WSNs constraints related to security issues.\\



\begin{enumerate}

\item Entity authentication

\indent An attack on a network can be extended to more than just modifying the data packets originally circulating in the network. An attacker can inject additional data packets to disturb the normal function of the network and tamper with the decision making process. For this reason, a receiver (i.e., node) must be sure that the data being accepted is coming from a member of the network. Similarly, a sender needs to verify that the reception entity is whom it claims to be. This finality can be achieved through authentication.\\
Benenson \textit{et al.} based their entity authentication on elliptic curve cryptography \cite{Benenson05}. Each user holds a legitimate certificate, which is the public key signed by a certification authority. Every node can verify the legitimacy of the users since the public key with the signature are preloaded in the sensors. Yet, this scheme requires an significant overhead for data encryption.\\

\item Backward and Forward Secrecy

\indent By capturing a sensor node, or inserting a malicious one, an adversary tries to gain access to confidential information in the network. In order to prevent this from happening, nodes should be forbidden from decrypting old messages (if the node joins the network) and future messages (if the node leaves the network).\\

\item Data confidentiality

\indent One of the most important issues related to network security is data confidentiality, and it refers to limiting data access to legitimate destinations. Keeping data packets confidential mainly means that:

\begin{itemize}

\item Sensor readings can only be performed by the legitimate destination; a sensor node holding information must not leak information to its neighbors.
\item Communication channel has to be secured, especially when the data being communicated is highly sensitive.
\item The network needs to achieve confidentiality by encrypting data during transmissions.

\end{itemize}

In \cite{Bahi}, Bahi \textit{et al.} argue that in-network communication, node scheduling, and data aggregation need to be proven as secure. For this matter, they proposed a security framework for wireless sensor networks. The authors proved that in-network communication answers to security objectives (indistinguishability, non-malleability, detection resistance). In addition to this, the proposed algorithm is able to aggregate data over encrypted packets.\\

\item Data freshness

\indent Data freshness means that data circulating in the network is recent and that no old messages are being replayed. In order to secure the network, shared keys need to be changed over time. During the update propagation time, an adversary can perform a replay attack, especially when the sensor node is unaware of key changing time.\\

\item{Data integrity}

\indent Data packets need to be maintained safe and unchanged over their life-cycle. Even harsh environment can take part in altering data while being routed. This is why it is crucial to implement mechanisms ensuring that, for a data packet, information being sent is equal to the information being received.\\

\item Secure localization

\indent Since there exists no physical connection between nodes in a WSN, it is highly important for the network to be able to accurately locate each sensor in the network. Yet, to maintain the network's integrity, the locations need to be recorded as secret information.\\



\item Time synchronization:

\indent As discussed previously, energy is an issue for WSNs. For this matter, sensor nodes need to go to sleep when they are neither sending nor receiving a data packet. Thus, the network needs to be time synchronized so the sensor nodes would be awake when they are needed in the transfer process.\\













\end{enumerate}

\end{enumerate}

\subsubsection{Defensive measures}

\indent Key establishment techniques have received great attention for many years. Nevertheless, WSN applications are relatively recent. Besides, the features of these networks are different from traditional networks. Therefore, preexisting techniques for key establishment are an unsuitable solution for WSNs applications. Traditionally, key exchange techniques use asymmetric cryptography (public key cryptography). Unfortunately, low power WSNs are unable to handle such a computationally intensive technique.\\
\indent The easiest way for encryption keys distribution, is to establish one single key for the entire network and forward it. It is easy to notice that this method is inefficient as one node can compromise the entire network.\\
\indent An alternative solution that can be adopted is symmetric encryption key. This technique secures communication between two hosts as they share a private key that is not recognized by the rest of the network. This key will be used for both data encryption and decryption.\\
\indent Another possibility is random probabilistic key distribution scheme. The initialization stage starts with preloading in every sensor node a maximum number of keys (with respect to the memory). This is done in a way that two sets of keys (in two different nodes) will at least share one key. By broadcasting the identity of the keys, every node can discover the neighbors with which it can exchange information. Now, every node can only communicate with its legitimate neighbors; a link only exists between nodes sharing a key. It is now possible for a sensor node to safely establish a link with a target node by secretly sharing a key via their neighbors \cite{Li11}.\\

\subsubsection{Means to reach dependability}

\indent In this section we will discuss different ways to increase the dependability of a network.

\begin{enumerate}

\item Fault prevention

\indent
\indent Once detected, it is important to prevent the fault from being incorporated into the network. Fault prevention starts with the design of the network (efficient design rules), through the implementation (simulations, structured programming), and during network operations (network maintenance, network protection).\\

\item Fault removal and forecasting

\indent Fault removal reduces the number and severity of faults within the network. It can be performed during the implementation of the network. The network design can be verified while being developed through verifications. Fault removal can also be performed while the network is being used. This is achieved through the maintenance cycle via corrective maintenance and preventive maintenance.\\
\indent As for fault forecasting, it estimates the present number of faults, their future incidence and the likely frequencies of their future occurrences. It aims at removing the effects of faults before their occurrences.\\

\item Fault tolerance
\label{tolerance}

\indent In order to prevent from interrupting the network operation, it is important to carry out mechanisms that will allow the network to continue delivering the required service even in the presence of active faults. Fault tolerance is a technique that allows the network to continue delivering correct service until full recovery, without any interruption \cite{Akyildiz02}.
\indent Geeta \textit{et al.} presented a fault tolerant communication framework for WSNs regarding the remaining energy in sensor nodes \cite{Geeta13}.\\
\indent Fault tolerance in wireless sensor networks is important for several reasons \cite{Koushanfar04}:

\begin{itemize}
\item \textbf{Technology and implementation aspects:} a WSN is exposed to interaction with its environment, causing hardware degradation. Plus, the network is required to perform a variety of actions under energy constraints.
\item \textbf{Complexity:} the complexity of the application will grow as the complexity of architecture and technologies increases. This will render the testing phase more and more complicated.
\item \textbf{Relatively recent scientific field:} research related to WSNs is still an open field, where there is no best way to address a problem, and no mistake-free solutions. 
\end{itemize}

\end{enumerate}

\section{Prognostics and Health Management: State of the Art}

\begin{figure}[!h]
\centering
\includegraphics[width=0.8\textwidth]{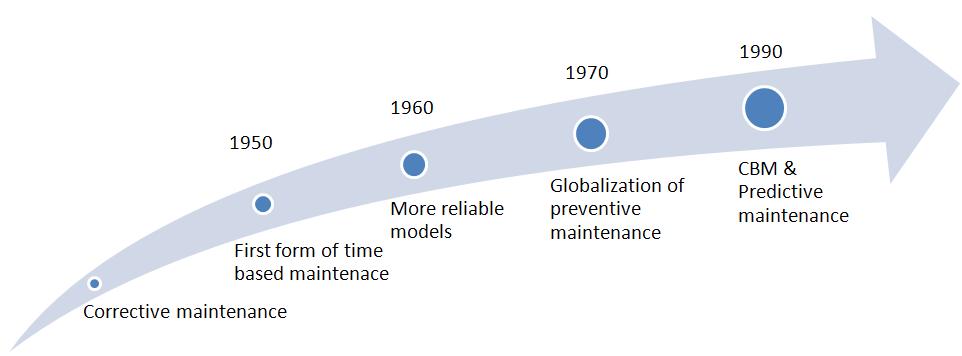}
\caption{History of maintenance strategies}
\label{his}
\end{figure}

\indent Maintenance is an important activity in industry. It is performed either to revive a machine/component, or to prevent it from breaking down. Different strategies have evolved through time, bringing maintenance to its current state. This evolution was due to the increasing demand of reliability in industry. Nowadays, plants are required to avoid shutdowns while offering both safety and reliability \cite{Peng10}.\\
\indent The first form of maintenance is corrective maintenance. In this strategy, actions are only taken when the system breaks and can no longer perform the intended tasks. Yet, plants cannot afford to undergo breakdowns; in fact, sudden shutdowns cost money and time, in addition to safety and clients' trust. As a remedy to this problem, maintenance became a periodic activity. Domain experts rely on their knowledge and the observation of upcoming events to set time intervals in which the components are inspected and replaced if needed. This preventive (often called periodic) maintenance is especially adopted by transportations and nuclear plants \cite{Hu12}. The main drawback of preventive maintenance is the fact that it is performed regardless of the machine's condition. In other words, industrials have to hire domain experts in order to set intervals for maintenance. Sometimes, this is unnecessary as the machine can be in a healthy state and this will cost extra and avoidable fees. Besides, even with periodic maintenance and inspections, random failures still occur. This is why Condition Based Maintenance (CBM) was proposed and developed in early nineties \cite{Heng09}.\\
\indent CBM is a proactive precess for maintenance scheduling, based on real-time observations. It is an online model that assesses machine's health through condition measurements. As any maintenance strategy, CBM aims at increasing the system reliability and availability. The benefits of this particular strategy include avoiding unnecessary maintenance tasks and costs, as well as not interrupting normal machine operations \cite{Heng09}.\\
\\
\indent Figure \ref{his} summarizes this evolution of maintenance strategies through time.

\begin{figure}[!h]
\centering
\includegraphics[width=0.5\textwidth,height=0.3\textheight]{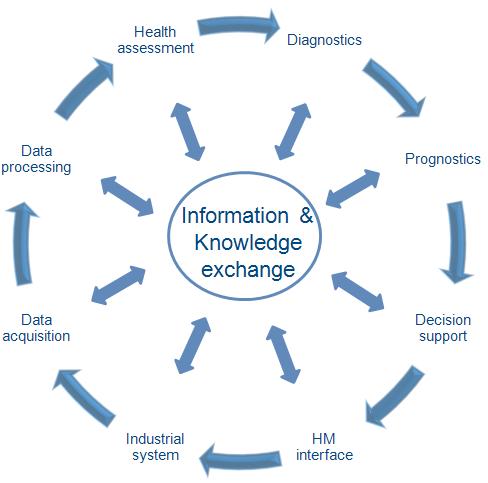}
\caption{CBM Flowchart}
\label{cbm}
\end{figure}

\indent In order to be efficient, a CBM program needs to go through the following steps \cite{Jardine06}, as illustrated in Figure \ref{cbm}.

\subsection{PHM: definitions}

\indent The terms diagnostics and prognostics are widely used. Though, the difference between these two concepts is sometimes vague. However, it is important to specify the difference as it is the key to perform a good PHM. \\
PHM is the core activity of CBM, and it implies the same steps. In the following, we will briefly discuss these steps, namely: data processing, health assessment, diagnostics, prognostics, and decision making support.

\subsubsection{Data acquisition and processing}
\label{acquisition}

\indent Remaining Useful Life (RUL) prediction requires information about the targeted physical assets. Such information can contain either event-data or Condition Monitoring (CM) data.\\
\indent Event-data reveal what was done, what happened, and what were the causes (repair, breakdown, installation...). In the other hand, CM data contain measurements related to the machine's condition. These two types of information are equally important for RUL prediction. Once this information is available, it is very important to perform data cleaning. This cleaning aims at isolating all the possible faults and avoiding the so-called \enquote{garbage in, garbage out} problem. Reported data can have a value type, a waveform type, or a multidimensional type. The two last types can contain noise and thus be very hard to exploit. Data processing is an important step as it converts data into useful information. Many processing techniques have been reported in prognostic literature \cite{Diego12,Niu10}, like wavelet decomposition, data denoising, data smoothing...

\subsubsection{Health assessment and diagnostics}

\begin{figure}[!h]
\centering
\includegraphics[width=0.65\textwidth,height=0.08\textheight]{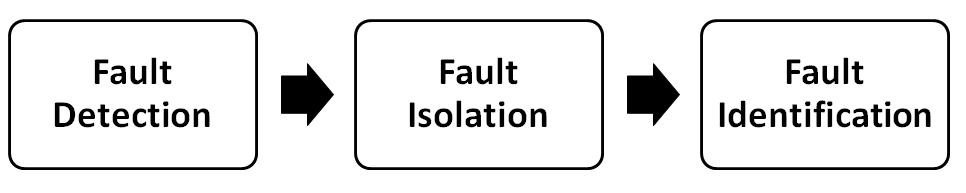}
\caption{Diagnostic's different steps}
\label{diagnostics}
\end{figure}

\indent Sensory data are reported periodically to monitor critical components. These data correspond to measurements of parameters (pressure, temperature, moisture...), and are useful to assess the machine's condition. Thresholds related to the monitored parameters are fixed. Once a threshold is reached, the system is considered to be in the corresponding state.\\
\indent Diagnostics is performed after the fault takes place. It aims at relating the cause to the effect. Diagnostics is an understanding of the relationship between what we observe and what happened before \cite{Sikorska11}.\\
In Figure \ref{diagnostics}, the successive steps of a diagnostic process are illustrated.

\begin{itemize}
\item Fault detection is used to report an anomaly in the system behavior.
\item Fault isolation is charged of determining and locating the cause (or source) of the problem. It identifies exactly which component is responsible of the failure.
\item Fault identification aims at determining the current failure mode and how fast it can spread.
\end{itemize}

\subsubsection{Prognostics}

\indent While diagnostics aims at identifying and quantifying an actual failure, prognostics have the goal of anticipating failures. Several definitions concerning prognostics exist in the literature. We summarized some of them in Table \ref{def}.

\begin{table}[!h]
\centering
\small
\begin{tabular}{|p{9cm}|c|c|}
\hline
Definition & Authors & Reference \\
\hline
\hline
&&\\
Estimation of time to failure and risk for one or more existing and future failure modes.& ISO $13381$-$1$ & \cite{ISO05}\\
&&\\
Estimation of the time before failure, or the remaining useful life, and the associated confidence value.& Tobon-Mejia \textit{et al.} & \cite{Diego12,Tob12}\\
&&\\
Indicates whether the structure, system, or component of interest can perform its function throughout its lifetime with reasonable assurance and, in case it cannot, to estimate the remaining useful life.& Zio and Di Maio & \cite{Zio10} \\
&&\\
Predicts how much time is left before a failure (or more) occurs, given the current machine condition and past operation profile. & K.S. Jardine \textit{et al.} & \cite{Jardine06} \\
&&\\
\hline
\end{tabular}
\caption{Some definitions of prognostics reported in the literature}
\label{def}
\end{table}

\begin{figure}[!h]
\centering
\includegraphics[width=0.6\textwidth,height=0.25\textheight]{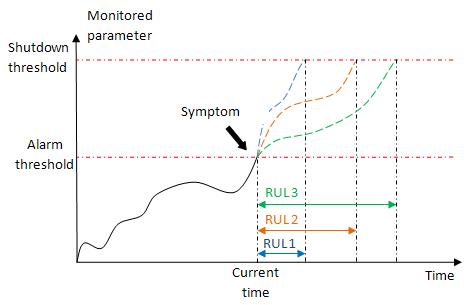}
\caption{An illustration of RUL with uncertainties}
\label{failure}
\end{figure}

\indent Prognostics considers past events, the machine's current state, and operating conditions to estimate the Remaining Useful Life (RUL). This estimation is done by inspecting the evolution of continuous measurements of parameters that need to be monitored in time to assess the machine's state. These parameters can be temperature, humidity, vibration, pressure, and so on. A monitored parameter has a fixed threshold. Once reached, an alarm goes off indicating that a symptom of system deteriorating has been detected. The RUL is then computed with an associated confidence limit. The latter information illustrates to what point the predictions are trustworthy. The uncertainties of the RUL predictions have two causes: either the threshold value of monitored parameter, or the RUL prediction itself.\\
In Figure \ref{failure}, we can observe the uncertainties that can be related to RUL prediction.

In \cite{ISO05}, the necessary pre-requisites for reliable prognostics are proposed.

\subsubsection{Decision support}

\indent Once prognostics are performed and RUL is estimated, the next step is to decide what are the actions that need to be taken (repair, replacement, maintenance, oil changing...). Decision making is a cognitive process. It consists of selecting an action among different possible scenarios, to produce a final choice.\\
\indent First of all, the objectives need to be established. An objective can be keeping component from failure until next inspection, reducing overall costs, or any other purpose a plant can be aimed for. All the objectives are then classified in order of priority and importance. Alternative actions are developed to answer the established objectives, and the actions that are able to satisfy most of the objectives are selected.\\
\indent A decision needs to be made to select the appropriate action. This decision can be made by implementing a tool among different possibilities.

\begin{itemize}
\item Domain experts: it is very often that plants trust the advices provided by engineers and domain experts. Thanks to their knowledge and experience, they are able to point out good solutions and uncover the limitations related to a strategy.
\item Eliminations: another solution is to eliminate non-realistic solutions one by one, or compare them in a pairwise manner. At the end, the remaining option is selected.
\item Analytic networks: these networks provide a hierarchy of the selected action with goals, alternatives, and consequences.
\item Simulations: there are many graphical tools used to visualize the behavior of a system under different conditions. Simulations are a popular tool for decision making support as they offer clarity and possibility to alter criteria while simulating.
\end{itemize}

\subsection{Classifying approaches}

\indent Prognostics approaches are classified under groups employing, more or less, the same techniques. Nevertheless, researchers use different classifications. Table \ref{class} summarizes the different groups we encountered during our study More details on each approach can be found in the given references.\\

\begin{table}[!h]
\centering
\small
\begin{tabular}{|l|c|c|}
\hline
Classifying groups & Authors & Reference \\
\hline
\hline
- Statistical approaches & K.S. Jardine \textit{et al.} & \cite{Jardine06} \\
- AI approaches & & \\
- Model-based approaches & & \\
\hline
- Event-based prediction & Heng \textit{et al.} & \cite{Heng09} \\
- Condition-based prediction & & \\
- Integrated approaches & & \\
\hline
- Physical model-based methodology & Peng \textit{et al.} & \cite{Peng10}\\
- Knowledge-based methodology & & \\
- Data-driven methodology & & \\
- Combination model & & \\
\hline
- Knowledge-based models & Sikorska \textit{et al.} & \cite{Sikorska11} \\
- Life expectancy models & & \\
- Artificial neural networks & & \\
- Physical models & & \\
\hline
- Model-based techniques & Cadini and Avram & \cite{Cadini09}\\
- Model-free methods & Zio and Di Maio & \cite {Zio10}\\
\hline
- Model-based approaches & Hu \textit{et al.} & \cite{Hu12} \\
- Data-driven approaches & & \\
- Hybrid approaches& & \\
\hline
- Model-based prognostics & Tobon-Mejia \textit{et al.} & \cite{Diego12} \\
- Data-driven prognostics & & \\
- Experience-based prognostics & & \\
\hline
\end{tabular}
\caption{Classifying models in the literature}
\label{class}
\end{table}

In this paper, we consider four groups: Physical models, Knowledge-based models, Data-driven models, and Hybrid models. They are detailed in the following sections.

\subsubsection{Physical models}

\begin{figure}[!h]
\centering
\includegraphics[width=0.8\textwidth, height=0.25\textheight]{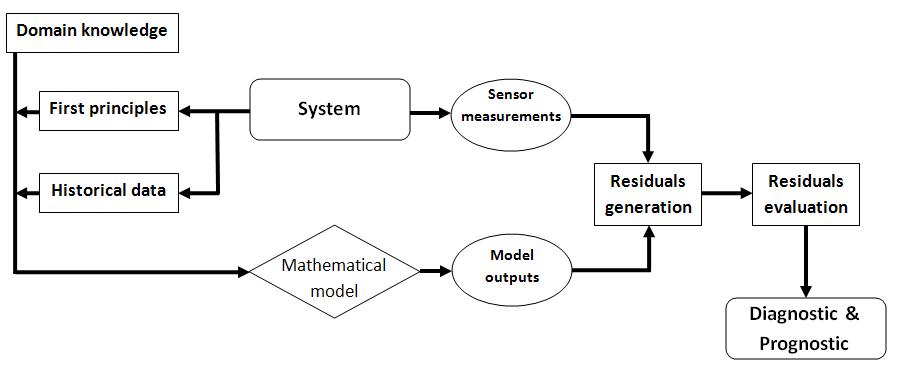}
\caption{Flowchart of a model-based approach}
\label{fizic}
\end{figure}

\indent Physical models rely on mathematical models to describe the physics of a failure, developed by domain experts. For this reason, the first condition for a reliable model is a good understanding of the behavior of the system responding to stress. The description of the behavioral models is realized via differential equations, state-space methods, or simulations.\\
In Figure \ref{fizic}, the general flowchart of a model-based approach is given.\\

Physical models are considered if:

\begin{itemize}
\item the mathematical model of the system is known;
\item the failure mode is well understood;
\item a physical model for each failure mode is available;
\item the operating conditions can be monitored; and
\item data describing the conditions related to each process is available.\\
\end{itemize}

\subsubsection{Knowledge-based models}

\indent Since it is really hard to build an accurate physical model for complex industrial systems, the employment of the latter is really limited. Besides, it is impossible to apply a developed model to a different component. Other methods, such as knowledge-based ones, appear to be promising as they require no physical model.\\
In the following, two examples of this model are presented.

\begin{enumerate}

\item Expert systems

\indent Since late 1960s, expert systems seemed to be suitable for problems usually solved by human specialists. These models consist of computer system, designed to display expert knowledge. This knowledge is extracted by domain specialists and organized into rules learned by the computer to generate solutions.\\
The general process of building such a model is described in Figure \ref{expert}.\\

\begin{figure}[!h]
\centering
\includegraphics[width=0.75\textwidth, height=0.08\textheight]{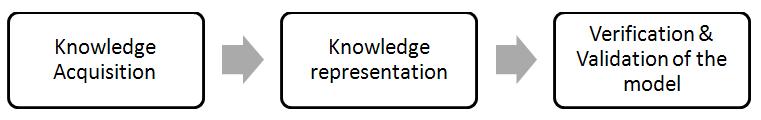}
\caption{Process of building an expert model}
\label{expert}
\end{figure}

The rules have the form of: 

\begin{center}
\textbf{IF} condition, \textbf{THEN} consequence.
\end{center}

Such a rule is strict and does not adapt to any changes in operating conditions. The only way to adapt the model to new situations is to add new rules whenever a new condition is observed. This can lead to a combinatorial explosion, especially that a rule is required for every possible combination of inputs. Another limitation of this model is that it is only as good as its developers.\\

\item Fuzzy logic

\indent Is a form of probabilistic knowledge, where the rules are approximate rather than fixed and exact. It was introduced by Lotfi A. Zadeh in $1965$ \cite{Zad65}.\\
\indent The difference between fuzzy logic and classical predicate logic, is the use of fuzzy sets rather than discrete values standing for true or false. In a fuzzy set, variable's membership is defined based on their degree of truth. The truth value ranges from $0$ (completely wrong) to $1$ (completely true).The rules may look like: 

\begin{center}
\textbf{IF} condition \enquote{A} \textbf{AND} condition \enquote{B} \textbf{THEN} consequence.
\end{center}

\indent The description associated to the parameters differs from the description used with expert system rules. Here is an example to illustrate the difference:

\begin{table}[!h]
\centering
\begin{tabular}{l l}
Expert system: & \textbf{IF} engine is hot \textbf{THEN} shutdown \\
&\\
Fuzzy logic: & \textbf{IF} engine is slightly hot \textbf{AND} temperature is rising \\ & \textbf{THEN} cool down  the system \\
\end{tabular}
\end{table}

\indent This new way of introducing rules gives the computer a very human-like and intuitive way of reasoning with incomplete, noisy, and inaccurate information. As a result, fault detection and prediction are more accurate, and for this reason, fuzzy logic is usually incorporated with other techniques.\\
\indent Even though this method can only be developed by domain experts, it is easy to understand the developed rules. It is not only recommended because  it covers a large set of operating conditions, but also because of its efficiency when it is impossible to build a mathematical model or when data contains high levels of uncertainties and noise.

\end{enumerate}

\subsubsection{Data-driven models}

\begin{figure}[!h]
\centering
\includegraphics[width=0.9\textwidth, height=0.26\textheight]{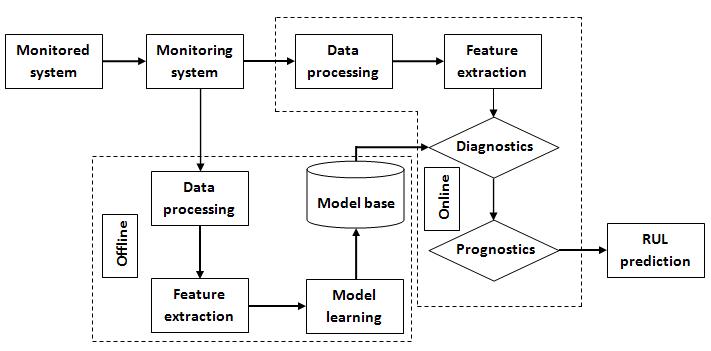}
\caption{General process of a data-driven approach}
\label{data}
\end{figure}

\indent In data-driven approaches, models are directly derived from condition monitoring data, based on statistical and learning techniques. These models have a double role: assess current operating conditions and predict the RUL. Neither human expertise nor comprehensive system physics are needed for the prognostic model building process.\\
\indent A data-driven prognostic model transforms raw data provided by the monitoring system into useful information (these techniques are discussed in Section \ref{acquisition}). By the means of this information and historical records, a behavioral model is built and predictions can be performed. The building process is detailed in Figure \ref{data}.

The data-driven approach is popular and widely used because it offers a tradeoff between complexity and precision. This approach remains the best solution when obtaining reliable sensor data is much easier than constructing mathematical behavioral models. Nevertheless, accuracy depends on many factors.

\begin{itemize}
\item The training set: normally, an efficient training requires a large set of inputs. It is not easy to decide whether the amount of inputs we dispose is enough for training a reliable model or not.
\item Operating conditions: manufacturing conditions change all the time, so do the environmental and operational conditions. All these changes may lead to uncertainties in the predictions as they refer to new situations that may not be recognizable by the model.
\item Sensory signals: the amount of effective sensory data available when prediction is performed has an impact on accuracy.
\item Degradation trend: RUL prediction relies on historical data and past events. As shown in Figure \ref{failure}, the prediction is an extrapolation of what we observe up to the present moment. If the degradation trend is highly similar to a trend the model can recognize, prediction can be accurate (and the other way around).
\end{itemize}

\indent Examples of the developed methods reported in the literature are:

\begin{itemize}

\item Aggregate reliability functions \cite{Crevecoeur93, Duane64, Goode00, Lee80, Noortwijk09, Todinov05}
\item Artificial neural networks ANN \cite{Huang07, Herzog09, Wang04, Brotherton00, Tsui95}
\item Autoregressive moving average ARMA \cite{Wu07, Yan04}
\item Bayesian techniques \cite{Cadini09, Kallen05, Weidl05, Weidl03, Ito00, Haug05, Marquez05}
\item Hidden markov and hidden semi-markov models \cite{Bunks00, Baruah05, Medjaher12, Dong07, He07}
\item Proportional hazards models \cite{Li07, Liao06, Makis03, Makis92}
\item Trend extrapolation \cite{Batko84, Kazmierczak83, Cempel87}

\end{itemize}

\subsubsection{Hybrid models}

\indent Usually, prognostic activity does not consider one parameter. The monitored parameters are diversified, as a consequence, it may be impossible to study failure behavior using only one model.\\
\indent Hybrid models aim at improving prediction quality by providing more accurate RUL. All research works agree that physical models guarantee the most precise prediction. Nevertheless, even with good outputs quality, the complexity is too important to ignore. This complexity can be reduced by adopting a data-driven approach. Thus, we can take benefits from the merits of both prognostic approaches.\\
When physical understanding of failure mechanism and monitoring data are available, a hybrid approach is the best solution offering a compromise between model complexity and prediction accuracy.\\
Table \ref{models} is a summary of each model's advantages and drawbacks.\\

\bottomcaption{An overview of Prognostic models}
\label{models}
\begin{centering}
\footnotesize
\begin{supertabular}{|p{3cm}|p{5.5cm}|p{5.5cm}|}
\hline
Prognostic Model & Advantages & Drawbacks \\
\hline
\hline
\textbf{Physical models} & - Require little data for prediction & - Very complex to build \\
& - Accurate and precise estimations & - Generates overall estimates\\
& - Can be reused in different conditions & - Needs complete knowledge of system behavior \\
& & - Component and defect specific\\
& & - Model validation requires a large set of data\\
\hline
\hline
\textbf{Knowledge-based models} &&\\
&&\\
Expert system & - Easy to develop and understand & - Hard to acquire domain knowledge\\
& - Mimics human thinking & - Accuracy requires many rules \\
&& - Hard to convert knowledge into rules\\
&& - Does not generate a confidence limit, nor accurate RUL \\
\hline
Fuzzy logic & - Handles imprecise and incomplete data & - Rules need to be developed by domain experts \\
& & - Decisions not easy to make \\
\hline
\hline
\textbf{Data-driven models} & & \\
&&\\
Aggregate reliability functions & - Simple and easy to understand & - Failures have to be statistically independent and equally distributed \\
&& - No warnings before failure \\
\hline
ANN & - Handles online pattern recognition & - Only efficient for a small set of data \\
& - Does not require any physical understanding of system behavior & - Needs retraining with every change of conditions \\
& - Models complex systems & - Needs pre-processing in order to reduce inputs \\
\hline
ARMA models & - Efficient for real-time applications & - Significant data for training \\
& - Does not need data related to failure history & - Does not benefit from prior knowledge\\
& - Does not require a detailed understanding of failure mechanism & \\
& - Accurate short term predictions & \\
\hline
Bayesian technique & - Does not require event data & - Relies on accurate thresholds\\
& - Predictions are easy to establish & - Needs a lot of state transitions for efficient prediction \\
& - Handles incomplete data & - Unable to predict unanticipated failures \\
&& - Prior knowledge needs to be available \\
\hline
HMM and HSMM & - Recognizes different failures & - Intensive computations \\
& - Able to recognize the change process & - The assumptions are not always practical \\
& - Models spatial and temporal data & - Large amount of data for training \\
& - Failure trend does not need to be monotonic & - Unable to predict unanticipated failures \\
& - Manages incomplete data sets & - State/Failure mode relation not clear \\
\hline
Kalman Filters & - Models multivariate and dynamic processes & - Can diverge easily\\
& - Handles incomplete and noisy data & \\
\hline
Particle Filters & - Provides non-linear projections & - Requires a large number of samples \\
&& - Not efficient for multi-dimensional data \\
\hline
PDF & - Accurate prediction close to failures & - Requires an important sample size \\
& - No need for CM data & \\
\hline
PHM & - Simple to develop & - Includes strict assumptions \\
& & - Requires historical data \\
\hline
Trend  & - Easy to apply & - Relies totally on past events \\
extrapolation & &- Needs a well-defined monotonic failure trend \\
\hline
\hline
\textbf{Hybrid models} & - Can be used with lack of historical data & - Requires both event and condition data \\
& - Accurate predictions & \\
\hline
\end{supertabular}

\end {centering}

\section{Wireless Sensor Networks for Industrial PHM}

\begin{figure}[!h]
\centering
\includegraphics[width=0.875\textwidth, height=0.22\textheight]{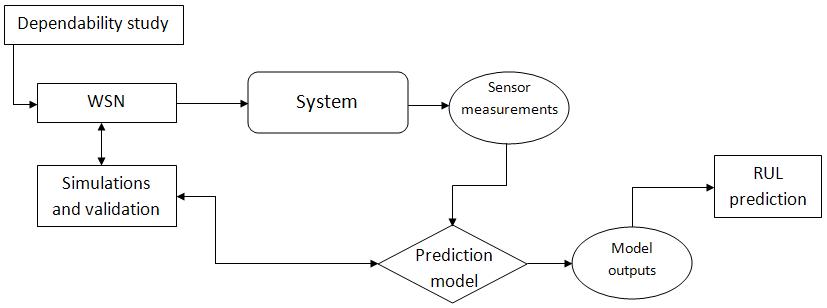}
\caption{General flowchart of CBM implementing a WSN}
\label{contr}
\end{figure}

\indent Reliability has become very essential in industry. It is a means to financial gain in addition to client trust. The research in the prognostic field, over the past years, resulted in a variety of tools and techniques offering plants the possibility to survey their systems, anticipate failures, and schedule maintenance. As the existent tools are different from one another, they have different advantages, drawbacks, complexities, etc. Data-driven prognostic models drew a great deal of attention due to their low cost, low complexity, and easy deployment. The prediction model will first acquire information about the monitored system, assess the current state, and then extrapolate the health state in the future. \\
\indent WSNs are mainly designed for surveillance purposes. They can be deployed in many fields such as military, automotive, agriculture, medicine...\cite{Li11}. Recently, industry has given WSN applications of monitoring a great deal of attention. Nowadays, they use sensor networks to monitor their machinery for maintenance scheduling. The sensors deployed to survey the system/component will provide data to estimate the RUL. Yet,  if this data is inaccurate, the prediction based on it will not be relevant. The dependability requirements, discussed before, need to be considered before the network starts running. Thereby, they can provide accurate data for RUL prediction and maintenance scheduling. Despite the existence of many dependability solutions in WSNs, these solutions are not always applicable. As sensors have restricted computational capabilities, solutions are often application-oriented. Thus, a definition of dependability issues related to prognostics is essential.\\
\indent As illustrated in Figure \ref{contr}, and before starting the predictions, a WSN dependability study needs to be taken into consideration. As good predictions rely on real data, it is obvious that the first step is ensuring a reliable source of information. Once the provided information are complete and correct, we will only need a robust prognostic model for good quality predictions.

\section{Challenges}

\indent Although many models have been developed for Prognostics and RUL prediction, there are many aspects that still need deep studying in order to provide more reliable predictions. How to use data fully? How to consider operating conditions in RUL prediction? How to allow multiple interactions while building a model? All these questions still need answers.\\
\indent Data-driven models are designed to reduce model complexity and enhance real-time maintenance. For this reason, they only provide general predictions for a population of identical units; this makes prediction process easier and faster.\\
\indent In the literature of prognostics, it is very common that the causes of a failure are limited to the values of monitored parameters. Other factors, although responsible of failures, seem to be neglected and overlooked. Although Condition Monitoring (CM) data reflect online monitoring, they do not replace reliability data. In fact, CM data provide measurements informing about a single component state at a specific moment. A failure does not only consider a single parameter (pressure, humidity...), it is a consequence of many factors (component age, different failing component...).\\ \indent Reliability data, informing about all these factors, give a bigger picture of the failing process. We are not neglecting the importance of CM data for prognostic process. Yet, while CM data provide information for short-term prediction, reliability data are able to extend these predictions until next maintenance window. The complete neglection of operating conditions, operating age, and interactions between failures can only limit the application of developed models to real machines. Operating conditions are never the same, they change all the time. If the model is unable to consider these changes, then it is unable to produce reliable estimations. Furthermore, if we observe two similar components with different operating ages and operating under similar conditions, we will notice that they will not fail at the same time. Operating age definitely has an influence on time to failure. Even a failure can accelerate or provoke another one. \\
\indent Another issue to face while performing prognostics, is censored data. Many plants do not allow their system to run to failure. Components are often replaced before they actually fail. As a result, the real time to failure is not kept record of. The performed preventive maintenance is mistaken for failure time, and RUL prediction is based upon that time. The value of RUL is critical for maintenance scheduling. In other words, the less accurate is the prediction, the less reliable is the maintenance schedule.\\
\indent Maintenance scheduling is the reason behind building prognostic models. Yet, once accomplished, the maintenance actions are not considered in the model. And generally, the related component is considered \enquote{as good as new}. It is very important to consider the effects of maintenance actions in the prediction model, at least to evaluate the model efficiency and study the new failure behavior after the maintenance being performed.\\
\indent What also drew our attention are the assumptions made to perform predictions. To the best of our knowledge, none of the previous research work has questioned the availability, safety, and security of data used for RUL prediction. It is assumed that:

\begin{itemize}
\item Sensory data is available and there is no data loss.
\item Sensor network is reliable.
\item There is no fault in sensors.
\item There is no constraint of energy consumption
\end{itemize}

So far, all prognostic work is limited to the condition monitoring layer, the health assessment layer, and the prognostic layer of the Open System Architecture for Condition-Based Maintenance OSA-CBM \cite{Thurston01, Niu10}. As RUL prediction concerns results that are yet to come, it has to rely on assumptions. Nevertheless, these assumptions, in no way, reflect a real life situation. The application of Wireless Sensor Networks (WSN) is very critical. First of all, the sensors' size is very small. So they have very small batteries with limited disposable energy. If the communication in the network does not consider this limitation, the sensors will quickly consume all the energy they have and be dropped. Thus, the information will no longer circulate in the network. Still, an energy efficient WSN will not stop some nodes from being dropped. This means that the network has to be fault tolerant in order to be able to pursue its functionalities in case of any sudden events (sensor loss, interferences...). Besides, like all wireless networks, WSN can be hacked. Competitors and hackers can steal information, change data, cause damage to the system... Data circulating in the network need to be secured against such attacks.\\
\indent Many research works have been done in WSN reliability field. But every application has its own features, and generalized solutions do not always solve the problem. An adapted solution for prognostics needs and goals should be considered.

\section{Conclusion}

\indent Condition-based maintenance is an important tool for modern plants in order to optimize their maintenance schedule. An appropriate schedule is reflected by the economical benefits. Choosing a prognostic model depends on key issues, such as model complexity, model strengths, and the amount of available information.\\
In the case where the industrial system is monitored by a wireless sensor network, data loss becomes highly probable, which has an important impact on the quality of predictions. In such a situation, the prognostic model is expected to maintain its robustness to the unpredictable lack of information.\\
Considering the actual challenges of real-life applications, condition-based maintenance needs to be oriented in a way to meet the expectations of modern plants. As existent research works suppose that data is complete, we believe that previous solutions are unsuitable for wireless sensor networks monitoring.\\
In future work, we intend to demonstrate the flaws of prognostic methods when data flow is non-continuous. We also intend to propose an algorithm that is able to provide good predictions even when data is incomplete.\\

\section*{Acknowledgments} We thank the anonymous referees for all the valuable comments that helped us to improve the quality of the paper.





\paragraph{\textbf{References}}

\bibliographystyle{elsarticle-num}
\bibliography{wiem}







\end{document}